\documentclass[twoside]{dis04}
\usepackage{epsfig}
\usepackage{graphicx}
\usepackage{bm}
\usepackage{amssymb,amsmath}

\def\runauthor{E. Iancu}
\def\shorttitle{The QCD $S$--matrix in the high--energy limit}
\newcommand{\be}{\begin{eqnarray}}
\newcommand{\ee}{\end{eqnarray}}

\def\simge{\mathrel{%
   \rlap{\raise 0.511ex \hbox{$>$}}{\lower 0.511ex \hbox{$\sim$}}}}
\def\simle{\mathrel{
   \rlap{\raise 0.511ex \hbox{$<$}}{\lower 0.511ex \hbox{$\sim$}}}}
\def\bigs{\mathrel{
   \rlap{\raise 0.531ex \hbox{$>$}}{\lower 0.531ex \hbox{$<$}}}}

\begin{document}

\title{The QCD $S$--matrix in the high--energy limit
}

\author{Edmond Iancu}

\address{Service de Physique Th\'eorique, CEA/DSM/SPhT,\\
91191 Gif-sur-Yvette Cedex, France\\
E-mail: eiancu@cea.fr }

\maketitle

\abstracts{I discuss elastic dipole--dipole scattering in QCD at high energies, 
with emphasis on the relation between Mueller's dipole picture
and the Color Glass Condensate, and on the importance of
rare fluctuations for the high--energy limit of the $S$--matrix.
}

\section{Dipole--dipole scattering: Color Dipoles versus Color Glass}

Let me consider the simplest scattering process that one can think
of in perturbative QCD --- the elastic scattering between two small 
color dipoles (or `onia') at zero impact parameter, 
in the center--of--mass (COM) frame and at relative
rapidity $Y$ --- and address a basic question: 
What is the high energy limit of the $S$--matrix for this 
collision ? On physical grounds, the answer seems quite clear:
$S_Y\to 0$ as $Y\to\infty$ (indeed, 
at sufficiently high energies, the  wavefunctions of the two dipoles 
contain so many gluons %---because of their `small--$x$' evolution---
that the probability $S^2_Y$ that no interaction take place in the
collision is very small, $S^2_Y\ll 1$, and it vanishes when $Y\to\infty$).
But deriving even this simple limit from perturbative 
QCD turns out to be quite non--trivial, as demonstrated, for instance,
by the `small--$x$ problem' of the linear evolution equations: 
The BFKL equation \cite{BFKL}, which is supposed to resum the dominant 
radiative corrections at high energies, predicts a scattering amplitude 
$T_Y\equiv 1-S_Y$ which rises exponentially with $Y$, thus eventually
violating the unitarity bound $T_Y\le 1$.

Over the last two decades, several formalisms have been
gradually developed which go beyond the BFKL evolution by including
those effects which restore unitarity at high energies. These effects
can be generically characterized as `multiple scatterings', but their
precise interpretation, and also their mathematical description,
depend crucially upon the choice of a frame: Whereas in the COM frame,
unitarity corrections start to manifest themselves, when increasing $Y$,
as genuine multiple scatterings (``multiple pomeron
exchanges'', or ``pomeron loops'') \cite{AM94}, in an asymmetric frame in which one
of the two dipoles is much faster than the other one, and thus carries
most of the evolution, unitarity also appears as non--linear
effects in the gluon distribution of the evolved dipole, or
``gluon saturation'' \cite{GLR,MV94}. 

This can be best appreciated on a simple example:
In the single--pomeron--exchange, or BFKL, approximation
(valid so long as $T_Y\ll 1$) and in the COM frame, 
the scattering amplitude 
for two identical dipoles of transverse size $r_0$ 
can be estimated as \cite{AM94}:
\be\label{1P} T(r_0,r_0,Y)\,\sim\,\alpha_s^2\,n^2(r_0,r_0,Y/2),\ee
where $n(r_0,r_0,Y/2)$ is the number density of 
radiated dipoles of size $r_0$ in the wavefunction of the 
parent dipole of size $r_0$ and with rapidity $Y/2$, and
$\alpha_s^2$ is the scattering amplitude for two elementary
dipoles of similar sizes. The BFKL evolution is encoded in
the dipole number density, which grows roughly like
$n(r_0,r_0,Y) \,\sim\,{\rm e}^{\omega_{\mathbb P} Y}$
with $\omega_{\mathbb P} = (4 \ln 2)\alpha_s N_c/\pi$ (the `BFKL intercept').
% and $\bar\alpha_s = \alpha_s N_c/\pi$.
Unitarity corrections become important when $T_Y\sim 1$, that is,
for $Y=Y_0$ with 
\be\label{Y0}
Y_0 \,\simeq\, {1\over \omega_{\mathbb P}} \ln \frac{1}{\alpha_s^2}\,.
\ee
For $Y\sim Y_0$, and in this particular frame,
BFKL fails to describe correctly the scattering amplitude, but it 
remains a reasonable approximation for the {\it wavefunctions} 
of the evolved dipoles. Indeed, the non--linear effects within
each wavefunction are rather measured by $\alpha_s^2 n(r_0,r_0,Y/2)$ 
--- the scattering amplitude between a given dipole and the other
dipoles within the same  wavefunction ---, which is still small,
of ${\mathcal O}(\alpha_s)$, when $Y\sim Y_0$. Therefore, saturation
effects in the COM frame start to manifest themselves only at higher
rapidities $Y\simge 2Y_0$. 

But when the same scattering is viewed in the asymmetric frame
in which one of the dipoles is nearly at rest, then
$T(r_0,r_0,Y)\,\sim\,\alpha_s^2\,n(r_0,r_0,Y)$, with $n(r_0,r_0,Y)$
referring to the evolved dipole. Again, $T$
becomes of ${\mathcal O}(1)$ for $Y\sim Y_0$, but when
this happens, the non--linear effects in the wavefunction of the
evolved dipole are also of ${\mathcal O}(1)$, and must be treated
on the same footing as the multiple scattering.

The language  of the example above is that of 
the {\it Color Dipole Picture} (CDP), a formalism originally developed 
by Mueller \cite{AM94} which cannot
accomodate saturation effects --- the onium wavefunction
is rather described by BFKL (together
with a large--$N_c$ approximation), as a collection
of dipoles which evolves through dipole splitting ---, but which can
describe unitarity corrections in the COM frame, as the multiple
scattering between several pairs of dipoles from the two incoming `onia'.
This formalism has been numerically implemented
by Salam \cite{Salam95}. It also lies at the basis of Kovchegov's derivation 
of a non--linear evolution equation for the scattering between a dipole 
and a large nucleus \cite{K}, to which I shall return later.

More recently, a different formalism has been developed, the {\it Color
Glass Condensate} (CGC) \cite{MV94,CGC}, which is specially tailored
to describe saturation in the wavefunction of an energetic hadron.
The CGC is the matter made of small--$x$ gluons in the high--density
environment characteristic of saturation. It is characterized by a
{\it saturation plateau} at relatively low transverse momenta :
the gluon modes with $k_\perp \le Q_s(Y)$ have large
occupation numbers, of ${\mathcal O}(1/\alpha_sN_c)$, but which
increase only slowly when increasing $y$ or decreasing $k_\perp$. The
{\it saturation momentum} $Q_s(Y)$   \cite{GLR} is an intrinsic
scale generated by the non--linear dynamics; it increases rapidly with 
$Y$, so at large $Y$ it provides a 
{\it hard} scale for the running of the coupling.

The mathematical
language of the CGC theory is that of {\it classical statistical physics} :
This is a theory for {classical color fields} endowed
with a (functional) probability distribution which evolves with $Y$
according to a (functional)
Fokker--Planck equation --- the JIMWLK equation \cite{CGC,JKLW97,W} ---
in which $Y$ plays the role of `time'.
The classical fields are generated by color sources
(the parent dipole plus radiated gluons) at rapidities larger than
the rapidity $Y$ of interest, whose dynamics is therefore
`frozen' by Lorentz time dilation. When increasing $Y$ in one more step
($Y\to Y+dY$ with $\alpha_s dY\sim 1$), a new layer is added to the
classical fields in longitudinal direction, corresponding to 
`integrating out' the emitted gluons with rapidities between $Y$
and $Y+dY$. The evolution thus generates a {\it random walk} in the 
configuration space of color fields, in which the elementary
step consists in the emission of a small--$x$ gluon 
in the background of the color fields created in the previous steps.
In this description,
saturation effects enter through the fact that the probability for induced
gluon radiation is non--linear in the background field and cannot exceed
one. The BFKL evolution is recovered in the limit where the  background 
field is weak, corresponding to low gluon occupation numbers.

Computing a scattering cross--section in the CGC formalism is a delicate task:
Since the color glass is characterized by strong fields, the standard 
factorization schemes for high--energy scattering are not bound to apply,
and in general we only know how to describe the collision between a CGC
and a simple projectile, like a (bare) dipole. Specifically, the
$S$--matrix for elastic CGC--dipole scattering is computed in the eikonal
approximation as ($x_\perp$ and $y_\perp$ are the transverse coordinates
of the quark and the antiquark which make up the dipole)
\be\label{Sdef}
S_Y(x_\perp,y_\perp)\,
%\equiv\frac{1}{N_c}
%\Big\langle {\rm tr}\big(U^\dagger(x_\perp) U(y_\perp)\big)
%\Big\rangle_Y\!= \!
=\int {\rm D}[\alpha]\,\, W_Y[\alpha]\,\,\frac{1}{N_c}\,
{\rm tr}\big(U^\dagger(x_\perp) U(y_\perp)\big)
,\ee
where $U^\dagger(x_\perp)$ and $U(y_\perp)$ are Wilson lines
describing the color precession of the quark, or the antiquark, in the 
color field $A^+_a\equiv \alpha_a$ of the CGC, 
%e.g., \begin{equation} \label{Vdef} U^\dagger(x_\perp)
%\,\equiv\,{\rm P}\,{\rm exp}\left({\rm i}g\int\! dx^-\,
%\alpha_a(x^-,x_\perp) t^a\right),
%\end{equation}
and $W_Y[\alpha]$ is the probability distribution for this field, which obeys the
JIMWLK equation alluded to above. By using the latter, one can derive an
evolution equation for $S_Y$; however, this is not a closed equation, but only
the first equation in an infinite hierarchy originally obtained by Balitsky
\cite{B}. Still, a closed equation for $S_Y$ can be obtained within
a mean field approximation (MFA): This is the non--linear equation originally
derived by Kovchegov \cite{K} (in a different physical context, though, namely 
for the scattering between a dipole and a large nucleus, where the MFA is better
under control), and which is generally dubbed as the Balitsky--Kovchegov
(BK) equation.

Returning to our original problem of the dipole--dipole scattering,
one sees that, within the CGC formalism, this scattering
is most simply described in an
{\it asymmetric} frame, in which one of the dipoles carries
most of the total rapidity and has evolved into a CGC, while the other 
dipole is rather slow and can be described as a bare
$q\bar q$ pair, without additional gluons. 
This feature complicates the comparison with
the CDP formalism, where the same problem is most
naturally formulated in the COM frame. Still, as shown recently
\cite{CGCDIP}
through analytic manipulations, the two formalisms {\it are} in fact equivalent,
within the range in $Y$ in which they are both supposed to 
apply. This equivalence is the first
point that I would like to slightly elaborate on in what follow. The other point
is the role of {\it rare fluctuations} in the
approach of the $S$--matrix towards the `black--body' limit $S=0$ \cite{FLUCT}.
(See also Ref. \cite{KL04} for a related analysis.)

\section{COM scattering between two color glasses}

The interesting rapidity range for comparing CDP to CGC
is $Y_0\lesssim Y < 2Y_0$ (cf. Eq.~(\ref{Y0})) : for $Y \ll Y_0$,
both formalisms reduce to the BFKL approximation (and thus are 
obviously equivalent), while for $Y \ge 2Y_0$, CDP fails to apply
because of saturation effects in the wavefunctions of the incoming dipoles
(in COM frame). The first step towards
establishing the equivalence consists in factorizing the $S$--matrix
for the elastic scattering between two color glasses
in the COM frame. The second step consists in showing that the JIMWLK
evolution of the wavefunction of a color dipole reduces to the 
corresponding BFKL evolution (as implemented in CDP)
in the weak field approximation and for large $N_c$.

Concerning the first step, the factorization proposed in \cite{CGCDIP}
reads as follows:
\be\label{SYCGC}
S_Y
\,=\,\int {\rm D}[\alpha_R]\, \,W_{Y/2}[\alpha_R]\int {\rm D}[\alpha_L]\, \,
W_{Y/2}[\alpha_L]
\,\,{\rm e}^{\,{\rm i}\int d^2z_\perp\,\nabla^i \alpha^a_L(z_\perp)
\nabla^i\alpha^a_R(z_\perp)}\,,
\ee
where the symbols $L$ and $R$ stand for the left--mover and the right--mover, 
respectively. The exponential factor is recognized
as the coupling between the color charge density in one system (e.g.,
$\rho^a_L=-\nabla^2 \alpha^a_L$) and the color field in the other system.
It describes multiple (eikonal) scattering in the
approximation that the {\it individual} color sources within each
system undergo at most {\it single} scattering. (But {\it global} multiple
scattering is still allowed, as the simultaneous scattering 
of several constituents from the two systems.) Eq.~(\ref{SYCGC})
is correct for $Y_0\lesssim Y < 2Y_0$ since in that range none of the two incoming
color glasses is at saturation, and multiple scattering of a {\it single}
gluon is indeed negligible. However, {\it global} multiple scattering 
{\it is} important, because each system involves a large number of
constituents. %(Recall the discussion around Eq.~(\ref{Y0}).)

As for the second step, one needs to show that the CGC and CDP descriptions
of an evolved dipole (the onium) become equivalent with each other
when the CGC formalism is simplified by using the weak field (or BFKL)
approximation together with the large--$N_c$ limit. To that
aim, we have shown in \cite{CGCDIP} that (i) the parent dipole can be
represented as a color glass, and (ii) its evolution with $Y$, as described by 
the correspondingly simplified version of the JIMWLK equation, can be 
reformulated as the evolution of a system of dipoles, in agreement with CDP.

The manipulations in Ref. \cite{CGCDIP} imply that Eq.~(\ref{SYCGC}) 
can be rewritten in the form expected in CDP  \cite{AM94}, namely, as the 
$S$--matrix for the scattering between two systems of dipoles. Schematically,
 \be\label{Sy}
S_Y=\sum_{N,N'=1}^\infty\int \! d\Gamma_N
P_N(Y/2)\int \! d\Gamma_{N'} P_{N'}(Y/2)\,{\rm exp}\bigg\{\!
- \sum_{i=1}^N\sum_{j=1}^{N'}\,T_0(i|j)\bigg\}\,,
\ee
where $P_N(Y/2)$ is the probability density for producing a given configuration
of $N$ dipoles after a rapidity evolution $Y/2$ (this depends
upon the transverse coordinates of the dipoles and evolves through
dipole splitting according to the BFKL kernel \cite{AM94,CGCDIP}), the integral
$d\Gamma_N$ runs over the dipole coordinates, and $T_0(i|j)$ is the
elementary scattering amplitude (via two gluon exchange) between
the dipole $i$ in the first onium and the dipole $j$ in the second one.

Note that, although the exponential in Eq.~(\ref{Sy}) looks formally like a 
Glauber approximation (the multiple scattering series is resummed
as the exponential of minus the amplitude for a single scattering),
this exponentiation holds only {\it configuration by configuration}. 
After averaging over all such configurations, the resulting $S$--matrix
differs significantly from the simple exponential of the one pomeron 
exchange\footnote{The one--pomeron--exchange ampolitude in
Eq.~(\ref{1P}) is recovered from
Eq.~(\ref{Sy}) as the linear term in the expansion of the exponential there.}.
This difference is particularly pronounced 
in the high energy regime at $Y\simge Y_0$,
where $S_Y$ is very small: The naive exponentiation of the one pomeron exchange
in Eq.~(\ref{1P}) would predict ($\kappa_0$ is an unknown factor):
\be\label{S1P}
S_Y \,\sim\,\exp\Big\{-\kappa_0 {\alpha}_s^2
n^2(r_0,r_0,Y/2)\Big\} \,\quad{\rm with}\quad
n^2(Y/2)\,\sim\,\,{\rm e}^{\omega_{\mathbb P} Y}\,,\ee
whereas the Monte--Carlo calculation of Eq.~(\ref{Sy}) by Salam rather yields
\cite{Salam95} :
\be\label{SSalam}
 S_Y \,\sim\,{\rm e}^{-\kappa\bar{\alpha}_s^2
Y^2}\,\quad{\rm with}
\quad \bar{\alpha}_s ={\alpha}_s N_c/\pi,\qquad
\kappa\approx 0.72\,,\ee
which, although small, is considerably larger than the naive estimate (\ref{S1P}).
Understanding this difference brings me to my next point, namely:

\section{On the importance of rare fluctuations at high energies}

Why is the $S$--matrix in Eq.~(\ref{SSalam}) approaching
the black--disk limit {\it so slowly} ? After all, Eq.~(\ref{S1P}) describes
the scattering between two {\it typical} configurations in the wavefunctions
of the incoming onia, by which I mean configurations which involve
a number of dipoles $N$ close to the average value $n(Y/2)
\sim\,{\rm e}^{\omega_{\mathbb P} Y/2}$ and for which $P_N = {\mathcal O}(1)$.
The huge difference between the estimates (\ref{SSalam}) and (\ref{S1P})
suggests that, at high energy, $S_Y$ is rather dominated by {\it rare}
configurations, which involve only few gluons ($N\ll n(Y/2)$), and therefore
have a very low probability to occur ($P_N(Y/2)\ll 1$), but which give a much larger
contribution to $S$ simply because systems with fewer dipoles
have a smaller probability to interact, and thus a larger probability $S^2$
to survive without interactions. (Note that the various configurations
contribute additively to the $S$--matrix in Eq.~(\ref{Sy}), so the 
sum there is dominated by those configurations which maximize the
product $P_N(Y/2)P_{N'}(Y/2)S_{N\times N'}$.)

At this point, it is interesting to note that a result very similar to that
in Eq.~(\ref{SSalam}) is obtained from the high--energy limit of the BK
equation \cite{LT99}, and also from approximate solutions to JIMWLK equation
valid deeply at saturation \cite{SAT}. Specifically, BK equation yields
\cite{LT99,FLUCT}
\begin{equation}\label{SBK}
S_Y(r_0) \,\simeq\,{\rm e}^{-{c\over 2}\bar{\alpha}_s^2(Y-Y_0)^2}\,,
\end{equation}
where $Y_0$ is such that $Q_s(Y_0) \sim 1/r_0$, and $c\approx 4.88$ is the 
exponent giving the energy dependence of the saturation momentum \cite{GLR}: 
$Q_s^2(Y)\simeq Q_s^2(0){\rm e}^{c\bar\alpha_s Y}$. Now, in the asymmetric frame
in which one dipole is bare and the other one is highly evolved, the BK equation
describes the scattering between the bare dipole and the {\it typical}
configuration in the evolved one, which is a CGC with saturation momentum
$Q_s(Y)$ \cite{SAT,FLUCT}. Then, the discrepancy between  Eqs.~(\ref{SBK}) and
(\ref{S1P}) shows that typical configurations play very different roles in
different frames, thus illustrating the strong sensitivity of the physical
picture of the high--energy scattering upon the choice of a frame. 
In particular, the configurations retained by
the BK equation in the COM frame must be some 
rare configurations, with only few gluons. But then it is
legitimate to rise doubts about Eq.~(\ref{SBK}) too :
Recall indeed that BK equation is obtained after a MFA, which should work
reasonably well for {\it typical} configurations, but not also for the {\it 
rare} ones ! But if rare configurations play such an important role at
high energies in the COM frame, there is no reason why they should be
less important in the asymmetric frame. In other terms, one cannot trust
any result like (\ref{S1P}) or (\ref{SBK}), which is
obtained by including the typical configurations {\it alone}.

Unfortunately, there seems to be no systematic way to identify
the rare configurations which dominate the $S$--matrix at high energies.
As a rough criterion, the relevant configurations must involve the {\it
maximal} number of gluons which can still give a contribution to
$S$ of order one: indeed, further increasing the number of gluons would rapidly
decrease $S$, whereas reducing the number of gluons even stronger would
suppress the probability of the configuration without significantly
enhancing its contribution to $S$. With this criterion in mind, we have been
able to `guess' some optimal configurations \cite{FLUCT}, and then 
check that they do indeed a
better job than the configurations retained by the BK equation, in the
sense of giving a larger contribution to $S$. Specifically, our best
configurations yield
\begin{equation}\label{SOPT}
S_Y(r_0) \,\simeq\,{\rm e}^{-{c\over 4}\bar{\alpha}_s^2(Y-Y_0)^2}\,,
\end{equation}
where as compared to Eq.~(\ref{SBK}) the exponent is now reduced by a factor
of two. (This result has been confirmed in \cite{KL04}.)
The optimal configurations look differently in different frames,
but they are rare in {\it any} frame. In particular, in the asymmetric
frame where typical configurations would lead to the result (\ref{SBK}),
the optimal, rare, configurations are those in which the fast onium has evolved
into a CGC having a lower than normal saturation momentum: $Q_s((Y+Y_0)/2)$
instead of $Q_s(Y)$.

To conclude, let me notice that the exponent in Eq.~(\ref{SOPT}) is still larger
than the one reported in the numerical calculation in Ref. \cite{Salam95} (cf.
Eq.~(\ref{SSalam})): $c/4\approx 1.22$ rather than $\kappa\approx 0.72$. So, it
would be worth redoing the numerical analysis in order to understand the
origin of this discrepancy.

\end{document}